\documentclass[sigplan,10pt,screen]{acmart}
\copyrightyear{2026}
\acmYear{2026}
\setcopyright{cc}
\setcctype{by-nc-nd}
\acmConference[SIGOPS ATC 2026]{The 2026 ACM SIGOPS Annual Technical Conference}{November 15--18, 2026}{Hong Kong, China}
\acmBooktitle{The 2026 ACM SIGOPS Annual Technical Conference (SIGOPS ATC 2026), November 15--18, 2026, Hong Kong, China}
\acmDOI{10.1145/3828523.3857132}
\acmISBN{979-8-4007-2818-1/2026/11}

\AtBeginDocument{%
  }

\usepackage{enumitem}
\newsavebox{\hybridqctablebox}
\newsavebox{\hybridqcfigurebox}

\graphicspath{{paper_sections/}}

\begin{document}
\pagestyle{empty}

\title{HybridQC: Hardware-Grounded Simulation of Tightly Integrated Hybrid Quantum-Classical Systems}
\author{Panayiotis Christou}
\email{pc33@fordham.edu}
\orcid{0000-0002-1172-3779}
\affiliation{%
  \institution{Fordham University}
  \city{New York City}
  \state{New York}
  \country{USA}
}

\author{Shuwen Kan}
\email{sk107@fordham.edu}
\affiliation{%
  \institution{Fordham University}
  \city{New York City}
  \state{New York}
  \country{USA}
}

\author{Ying Mao}
\email{ymao41@fordham.edu}
\affiliation{%
  \institution{Fordham University}
  \city{New York City}
  \state{New York}
  \country{USA}
}

\renewcommand{\shortauthors}{Christou et al.}

\begin{abstract}

{\leavevmode Hybrid quantum--classical performance depends on how quantum processors share classical compute, controllers, memory, and communication links. Circuit simulators and provider measurements expose kernel and job behavior, but leave topology-level questions about controller bottlenecks, quantum capacity, and mixed-workload contention unanswered. We present HybridQC, a discrete-event simulator for configurable hybrid compute units (HCUs). Users specify resource graphs containing classical compute, memory, controllers, quantum annealing (QA) and digital quantum computing (DQC) devices, and links. Jobs become dependent stages for preparation, quantum execution, transfers, reconstruction, and postprocessing. Finite capacities and selectable scheduling policies determine where stages wait and overlap. Service profiles use measurements from D-Wave Advantage 1 and Advantage 2 and IBM Kingston, Marrakesh, and Fez, separating physical quantum-processor occupancy from cloud wall time. Mean absolute percentage errors are 3.92\%--8.04\% for D-Wave access time and 5.26\%--19.01\% for IBM quantum seconds. Balanced 10$\times$ HCU scaling improves makespan by 2.19$\times$--3.42$\times$, while policy choice changes makespan by 1.80$\times$ for a mixed workload with 20 jobs per workload family. Scaling depends on which inputs grow: 100$\times$ data growth yields median runtime near 306 s, while 100$\times$ joint growth in circuits, shots, and depth reaches $4.806\times10^7$ s on the fixed HCU. HybridQC enables designers to study topology, scheduling, and workload growth before tightly integrated quantum--classical machines exist physically.}

\end{abstract}



\maketitle

\section{Introduction}
\label{sec:introduction}



Hybrid quantum-classical computing is rapidly shifting from loosely coupled cloud access toward tightly integrated accelerator deployments. A useful hybrid workflow is not a single call to a quantum processing unit (QPU), {\leavevmode a physical quantum processor used alongside classical processors}. It is a complex pipeline of classical preprocessing, problem mapping, embedding or transpilation, data movement, quantum annealing (QA) or digital quantum circuit (DQC) execution, result reconstruction, and postprocessing. Industry efforts such as NVIDIA NVQLink, EuroQCS, and HPCQS point toward deploying QPUs as local accelerators alongside CPUs and GPUs within the same compute node~\cite{nvidia_nvqlink,nvidia_nvqlink_blog,eurohpc_euroqcs,eurohpc_hpcqs}. End-to-end performance in these Hybrid Compute Units (HCUs) is therefore determined by how every pipeline stage contends for finite resources across the entire hardware topology.

{\leavevmode QA searches for low-energy solutions to an encoded optimization problem. DQC executes circuits of quantum gates. Embedding maps a problem onto an annealer's qubits and connections. Transpilation maps a circuit onto supported gates and connections.}



Existing tools cannot answer the architecture and system questions that matter for designing future tightly integrated HCUs. How many controllers are needed per QPU before controllers rather than QPUs become the throughput bottleneck? When does adding quantum capacity stop improving end-to-end makespan because a transfer link or classical reconstruction stage is already saturated? How do mixed QA and DQC workloads reshape resource contention and shift bottlenecks as job composition changes?


We present HybridQC, a hardware-grounded discrete-event simulator for configurable hybrid compute units that directly addresses these gaps. HybridQC is built around three design decisions that enable topology-level reasoning. First, an HCU is represented as a directed resource graph where CPUs, GPUs, QA devices, DQC devices, controllers, memories, and interconnects are all modeled as finite-capacity first-class resources. Treating controllers and links as explicit architectural resources rather than invisible wires is what enables HybridQC to distinguish a QPU bottleneck from a controller bottleneck. Second, hybrid workloads are lowered into typed stage directed acyclic graphs (DAGs) that make every pipeline dependency explicit. Problem preparation, embedding or transpilation, controller programming, quantum execution, readout, transfer, reconstruction, and postprocessing each appear as separate stages with their own resource requirements, runtime models, memory footprints, and communication payloads. {\leavevmode Compilation, circuit cutting, and quantum error mitigation (QEM) can be added as parameterized stages. QEM uses extra measurements or classical processing to reduce the effect of noise on estimated results.} Third, physical QPU occupancy is separated from cloud provider overhead at the modeling level. D-Wave QPU access time and IBM-reported quantum seconds represent true device occupancy and drive integrated-HCU resource modeling. Provider queue time, orchestration latency, and client delays are retained as a separate deployment mode. This separation allows today's cloud measurements to ground studies of future integrated systems without inheriting provider overhead as a device constant.

{\leavevmode A workload family is an application template from which jobs with different input parameters are generated.}

Validation against today's IBM and D-Wave hardware makes the simulator
credible, while the HCU model lets us forecast bottlenecks in tightly
integrated systems that do not yet exist. Across calibrated hardware backends,
diverse workload families, and multiple topologies and policies, this paper
makes 4 key contributions:

\begin{itemize}
    \item {\bf Hardware-grounded runtime separation}. We calibrate backend-specific profiles with 3.92\%--8.04\% {\leavevmode mean absolute percentage error (MAPE)} for D-Wave QPU access time and 5.26\%--19.01\% MAPE for IBM quantum seconds, using D-Wave SampleSets and IBM Kingston, Marrakesh, and Fez campaigns. We separate physical QPU occupancy from cloud provider overhead so the same data can support both present-day validation and future integrated-HCU forecasting.

    \item {\bf Non-linear topology scaling and persistent bottlenecks}. Balanced 10$\times$ hardware scaling improves makespan by only 2.19$\times$--3.42$\times$ rather than 10$\times$, because controllers, transfers, and classical stages remain on the critical path. Pair sweeps show why: QA-controller replication raises average throughput from 2.85 to 9.75 jobs/s, while DQC-controller replication reduces p95 completion time from 7601 s to 6348 s.

    \item {\bf Scheduling policies as first-class architectural parameters}. Queue management policy changes mixed-workload makespan by up to 1.80$\times$, showing that scheduler selection must be co-designed with topology and workload composition. {\leavevmode First-in, first-out (FIFO)}, Hybrid Overhead-Balanced Adaptive (HOBA), Greedy Max Pressure (GMP), Adaptive Batch-Optimal (ABO), and Max-Pressure expose different bottleneck behavior because they use queue state, batching, feed urgency, and pressure differently.

    \item {\bf Axis-dependent scaling cliffs}. We scale data, shots, circuit count, depth, and all workload fields separately across QA-only, DQC-only, and QA+DQC workloads. 100$\times$ data-only scaling produces manageable transfer pressure at a median of 306 s, while 100$\times$ joint growth in circuit count, shots, and depth reaches about $4.806\times10^7$ s on the fixed HCU and $1.280\times10^7$ s on the 10$\times$ larger HCU. These future-system stress tests reach hundreds to thousands of logical qubits and depths up to 200,000, beyond today's evaluated devices.

\end{itemize}

\section{Background and Related Work}
\label{sec:background}

\begin{figure*}[t]
  \centering
  \includegraphics[width=0.95\textwidth]{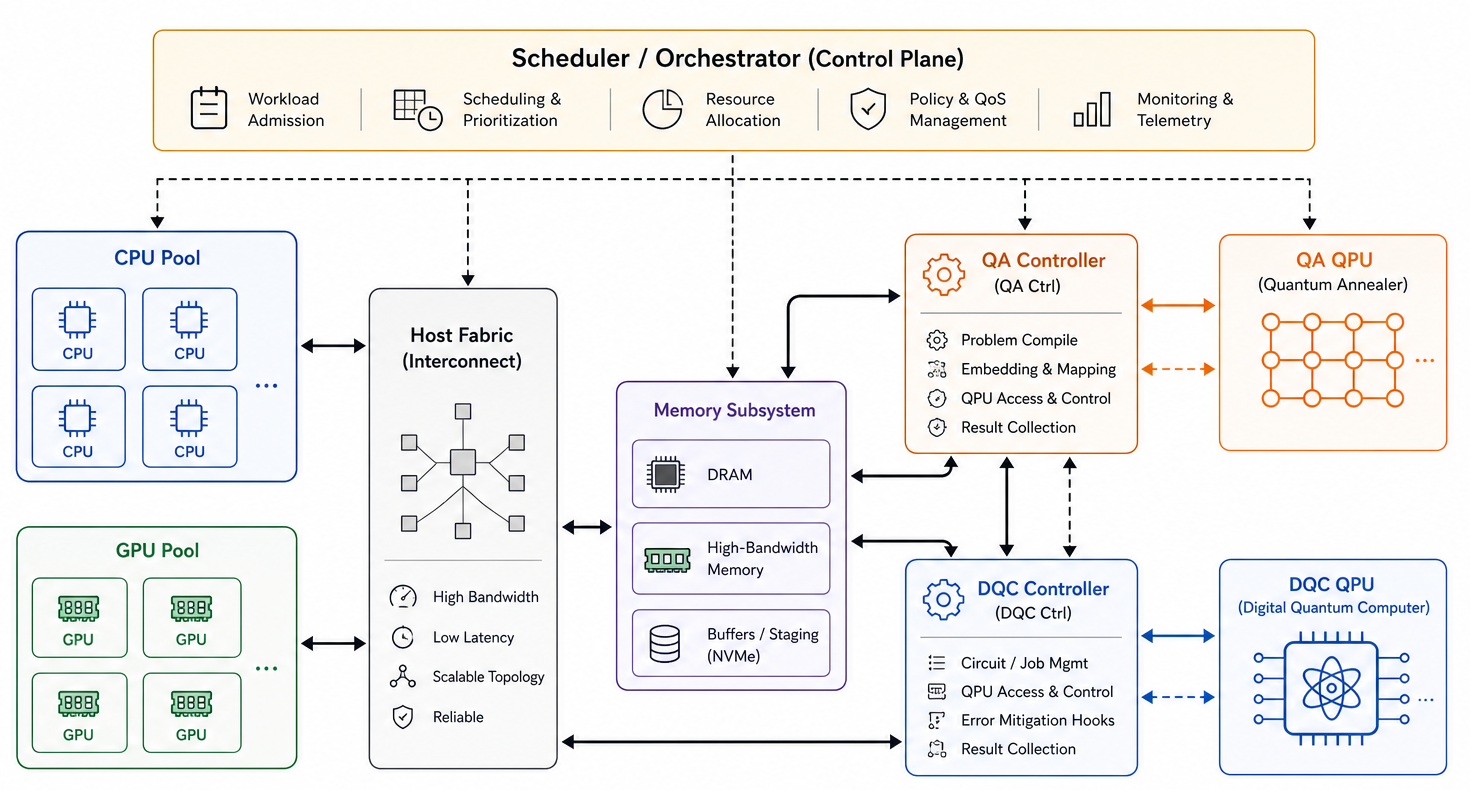}
  \caption{Configurable HCU topology modeled by HybridQC. Users define an HCU as
  classical and quantum compute nodes connected by explicit data/control links.
  HybridQC~then simulates stage-by-stage data flow, resource utilization, and
  contention across that topology. {\leavevmode Solid arrows show data paths; dashed arrows show control and orchestration connections. Orange and blue distinguish the QA and DQC paths. The controller cross-link is an optional topology edge.}}
  \label{fig:hcu_topology}
  \Description{A scheduler and orchestrator manages CPU and GPU pools, a host interconnect, memory, and separate quantum-annealing and digital-quantum controller and processor pairs. The host interconnect connects classical compute and memory to both controllers. Each controller connects to its corresponding quantum processor. An optional link connects the two controllers. Solid paths carry data, while dashed paths carry control and orchestration messages.}
\end{figure*}

\subsection{Hybrid Workloads Are Pipelines, Not Kernels}
\label{sec:bg-workloads}

{\leavevmode Applications for noisy intermediate-scale quantum (NISQ) computers are hybrid by construction.} {\leavevmode The variational quantum eigensolver (VQE) and quantum approximate optimization algorithm (QAOA) alternate classical updates with quantum evaluations}~\cite{peruzzo2014vqe,farhi2014qaoa}. Annealing workloads map
optimization problems into Ising-like forms~\cite{lucas2014ising}. NISQ
systems require compilation, sampling, mitigation, and postprocessing around
every device call~\cite{preskill2018nisq}. {\leavevmode A shot is one circuit execution and measurement. An annealing read returns one sample from the encoded optimization problem.} Recent sequential and hybrid
sequential quantum-computing proposals make this pipeline structure
explicit~\cite{romero2025sequentialquantumcomputing,
chandarana2025hybridsequentialquantumcomputing,kipuquantumSequentialQuantum}.
Their reported QA-plus-DQC hardware demonstrations make this more than a
modeling abstraction: coupling annealing search with digital refinement is
emerging as an industry-relevant hybrid execution pattern.

The workloads studied in this paper follow the same view. Portfolio,
bin-packing, unit-commitment, vehicle-routing, pipeline-optimization,
resource-allocation, {\leavevmode potential energy surface (PES) scans}, and VQE-style workflows each contain a quantum
kernel, but their end-to-end behavior is shaped by the stages around that
kernel: input construction, embedding or transpilation, transfer, QPU
execution, result collection, mitigation, and
reconstruction~\cite{Morapakula_2025,de_Andoin_2022,
sawamura2025quantumclassicalhybridalgorithmusing,
ellinas2024hybridquantumclassicalalgorithmmixedinteger,
maciejunes2025solvinglargescalevehiclerouting,
christeson2025hybridquantumclassicaloptimizationresource,pipeline_optimization,
Brown_2024}. \emph{End-to-end performance is therefore a property of the
pipeline, not of the kernel.}

\subsection{Integration is the New Design Target}
\label{sec:bg-integration}

Vendor and infrastructure efforts are moving away from cloud-attached QPUs
and toward tightly coupled hybrid compute units. NVIDIA NVQLink targets
sub-microsecond GPU--QPU coupling for accelerator-style
integration~\cite{nvidia_nvqlink,nvidia_nvqlink_blog,
nvidia-connectx6vpiocp3-intro}. The European EuroQCS and HPCQS initiatives
co-locate QPUs with HPC systems and expose them through HPC schedulers and
queues~\cite{eurohpc_euroqcs,eurohpc_hpcqs}. Recent measurement and
architecture studies in turn identify data movement, orchestration, queueing,
resource allocation, and accelerator scheduling as the dominant integration
challenges in such systems~\cite{doler2025surveyintegratingquantumcomputers,
scaling_hyrbid_qhpc,Beck_2024,rocco2025dynamicsolutionshybridquantumhpc,
viviani2025assessingelephantroomscheduling,tang2025multipathtransferenginebreaking}.

\subsection{The Gap: No Tool Simulates the Entire Quantum-Classical Topology}
\label{sec:bg-gap}

Existing tools capture important pieces of the stack, but none covers the full
topology of an integrated HCU. Circuit simulators model quantum kernels,
cloud and provider tools expose QPU jobs, classical simulators model CPUs,
memory, and networks, and quantum-network simulators model distributed quantum
communication.
Table~\ref{tab:related-work} summarizes the resulting gap. \begin{table*}[t]
\centering
\footnotesize
\setlength{\tabcolsep}{3pt}
\renewcommand{\arraystretch}{1.12}
\begin{lrbox}{\hybridqctablebox}
\begin{tabular}{@{}llp{0.16\textwidth}p{0.18\textwidth}p{0.12\textwidth}p{0.12\textwidth}@{}}
\toprule
\textbf{System} &
\textbf{Models QPU service?} &
\textbf{Models classical resources?} &
\textbf{Models controllers as first-class?} &
\textbf{Hardware-grounded?} &
\textbf{Topology-configurable?} \\
\midrule
\multicolumn{6}{@{}l}{\underline{\textit{Circuit simulators}}} \\
\addlinespace[0.15em]
Qiskit Aer / QuEST / Cirq / ProjectQ~\cite{qiskit_aer_docs,quest,cirq_zenodo,projectq}
& Yes, circuit-level & No & No & N/A & No \\
\addlinespace[0.25em]
\multicolumn{6}{@{}l}{\underline{\textit{System simulators}}} \\
\addlinespace[0.15em]
iQuantum / QSimPy~\cite{iquantum,qsimpy}
& Partial, service-level & Yes, cloud-level & No & Limited & Limited \\
\addlinespace[0.25em]
HyperQ~\cite{hyperq_osdi2025}
& Yes, virtualized QPU service & Yes & Partial & No & Yes \\
\addlinespace[0.25em]
gem5 / SST/macro / CloudSim~\cite{gem5_binkert2011,sstmacro2013workflow,cloudsim}
& No & Yes & N/A & Yes & Yes \\
\addlinespace[0.25em]
\multicolumn{6}{@{}l}{\underline{\textit{Network simulators}}} \\
\addlinespace[0.15em]
NetSquid / SeQUeNCe / QuISP~\cite{netsquid,sequence,quisp}
& Network-level & No & No & No & No \\
\midrule
\textbf{HybridQC~(this work)}
& \textbf{Yes, calibrated multi-stage}
& \textbf{Yes}
& \textbf{Yes}
& \textbf{Yes}
& \textbf{Yes} \\
\bottomrule
\end{tabular}
\end{lrbox}%
\ifdim\wd\hybridqctablebox>\textwidth
  \resizebox{\textwidth}{!}{\usebox{\hybridqctablebox}}%
\else
  \usebox{\hybridqctablebox}%
\fi
\caption{Positioning of HybridQC~relative to representative quantum and systems
simulators. The missing capability is not circuit simulation alone, but
hardware-grounded simulation of complete hybrid topologies with classical
resources, controllers, links, queues, and QPU service on the same time axis.}
\label{tab:related-work}
\Description{Rows compare circuit, system, and network simulators across quantum service, classical resources, controller modeling, hardware grounding, and topology configuration. The final row gives the capabilities of HybridQC.}
\end{table*}

\emph{Circuit simulators and provider runtime APIs} such as Qiskit, Cirq,
PennyLane, ProjectQ, QuEST, Quantum-Train, and Qiskit Aer are essential for
circuit construction, transpilation, and algorithm
development~\cite{qiskit_zenodo,qiskit_aer_docs,ibm_qiskit_serverless,
cirq_zenodo,projectq,quest,liu2024introduction,pennylaneQuantumProgramming}.
IBM and D-Wave provider metadata are likewise
indispensable for empirical grounding~\cite{ibm_estimate_runtime,ibm_sessions,
ibm-quantum-qpu-information,dwave_qpu_properties,dwave-annealing}. None of
these tools, however, models how a quantum call
interacts with the surrounding CPUs, GPUs, controllers, memories, links, and
schedulers of an integrated node.

\emph{Classical and cloud simulators} go further. gem5 and SST/macro model
classical architectures and coarse-grained HPC workflows, while CloudSim
models cloud-level resource provisioning~\cite{gem5_binkert2011,
sstmacro2013workflow,cloudsim}.
{\leavevmode iQuantum and QSimPy provide service-level models for quantum-cloud scheduling, while HyperQ targets QPU virtualization~\cite{iquantum,qsimpy,hyperq_osdi2025}. iQuantum is extensible: its classical-resource and quantum-task abstractions provide a starting point for hybrid scheduling studies. HybridQC decomposes applications into dependent classical, transfer, control, and quantum stages. Each stage reserves its required resources, exposing controller and link contention within an HCU. The model separates measured physical QPU occupancy from provider turnaround. These mechanisms connect workload dependencies to topology and service profiles.}

\emph{Quantum-network simulators} such as NetSquid, SeQUeNCe, QuISP, and
SimulaQron target distributed quantum
communication~\cite{netsquid,sequence,quisp,simulaqron}. They solve a
different problem and do not model intra-node hybrid compute.

HybridQC~combines these capabilities in a topology-level discrete-event simulator of tightly integrated HCUs.

\subsection{Controllers and Links as First-Class Resources}
\label{sec:bg-controllers}

The capability gap above has a concrete cause. Quantum systems depend on
substantial classical control infrastructure --- control electronics, host
interfaces, {\leavevmode field-programmable gate array (FPGA) controllers}, scheduling logic, and near-device
orchestration~\cite{classsical_control_electronics,
classical_interfaces_control,quantum_control_fpga,controller_executing_qec,
Shehata_2026}. These components are largely invisible in algorithm-level
benchmarks, yet they become architectural bottlenecks in tightly integrated
systems. The next section defines this resource graph and the configuration interface
that exposes it to the user.

\section{HybridQC: HCU Abstraction}
\label{sec:model}

\subsection{Layered Architecture and Inputs}
Figure~\ref{fig:hcu_topology} is the organizing object for HybridQC. An HCU is a
configurable topology of compute nodes---classical CPUs/GPUs, memories,
controllers, QA devices, DQC devices, and optional user-defined accelerators---
connected by explicit data and control links. Workloads enter at the
application layer as hybrid quantum-classical programs with per-job parameters,
data payloads, and stage dependencies. The workload layer lowers those programs
into multi-stage execution graphs. The runtime-profile layer attaches
backend-specific QA and DQC service models. The topology layer determines where
each stage can run and how data moves between nodes. Finally, the simulation
layer executes the resulting stages over the topology and records resource
utilization, data movement, queueing, wait, bottleneck, throughput, and
completion metrics.

This layering is {\leavevmode a central design choice}: HybridQC~models an HCU as a
user-configurable topology rather than as a fixed backend template or a single
QPU-runtime estimator. A user can change the topology without changing the
workload definition, change a scheduling policy without changing backend
profiles, or re-ground a backend profile with new hardware data without
rewriting the execution engine. The four main input classes are summarized in
Table~\ref{tab:hcu_inputs}.

\begin{table}[t]
  \centering
  \small
  \begin{tabular}{@{}p{0.25\linewidth}p{0.67\linewidth}@{}}
    \toprule
    Input class & Role and configurable examples \\
    \midrule
    Workload &
    Defines the multi-stage workload graph and data movement. Examples:
    workload family, shots, circuits, depth, QA/DQC payloads, and scaling axes. \\
    \addlinespace[0.25em]
    Topology &
    Defines what stages can run and where contention occurs. Examples:
    resources, capacities, links, memory, controllers, and user-added nodes. \\
    \addlinespace[0.25em]
    Control semantics &
    Defines programming, readout, mediation, and provider overhead. Examples:
    controller profile, routing mode, cross-device handoff mode, and cloud
    mode. \\
    \addlinespace[0.25em]
    Execution config &
    Defines queue ordering, experiment scope, and feasibility checks. Examples:
    policy, arrivals, workload scale, horizon, and capacity semantics. \\
    \bottomrule
  \end{tabular}
  \caption{Configuration inputs to HybridQC. The same workload can be evaluated
  across many HCU topologies, policies, and backend profiles.}
  \label{tab:hcu_inputs}
\Description{Four rows group the inputs into workload, topology, control semantics, and execution configuration. Each row explains its role and lists configurable examples.}
\end{table}

\subsection{Configurable Resource Graph}
HybridQC~separates the hardware object from the simulator that executes on it. A
hybrid compute unit (HCU) is a directed resource graph $G=(V,E)$. Nodes
represent finite resources: CPU pools, GPU pools, memories, QA devices, DQC
devices, QA controllers, DQC controllers, and user-defined accelerator or
control nodes. We use the latter term for resources such as specialized
controller processors, FPGAs, future QPU-specific mediators, or other
accelerators that expose capacity and service-time fields but are not one of
the default CPU/GPU/QA/DQC classes. Edges represent explicit data or control
paths with latency, bandwidth, and routing constraints. A topology therefore
specifies not only which QPUs exist, but how work reaches them and which
controller resources serialize the path.

Each resource node has a type, a capacity, and optional hardware limits. For
classical nodes, these include memory capacity and device count. For QA and DQC
nodes, they include problem-size limits such as available qubits,
coupler/logical capacity, or backend-specific circuit limits. {\leavevmode Table~\ref{tab:hcu_parameters_exact_app} gives the baseline timing and transfer parameters.} Section~\ref{sec:grounding} explains
how the QA/DQC timing terms are refined with backend-specific measurements.

\begin{table*}[t]
\centering
\caption{Baseline HCU parameters used by the simulator before backend-specific
calibration. The current evaluation refines QA and DQC service models using
measured D-Wave and IBM data, while retaining these fields as explicit
configuration dimensions.}
\label{tab:hcu_parameters_exact_app}
\Description{The table groups baseline settings into interconnect, quantum annealing, digital quantum computing, and data movement and constraints. Each parameter is paired with a value, range, timing formula, or resource rule.}
\setlength{\tabcolsep}{5pt}
\renewcommand{\arraystretch}{1.05}
\footnotesize
\begin{tabular}{@{}p{0.24\textwidth}p{0.68\textwidth}@{}}
\toprule
\textbf{Parameter} & \textbf{Value used or modeled} \\
\midrule
\multicolumn{2}{@{}l}{\textbf{Interconnect between classical host and QPU/control path}}\\
GPU--QPU bandwidth & 400 Gb/s, approximately 50 GB/s \cite{nvidia_nvqlink} \\
GPU--QPU latency & mean 3.84 $\mu$s, max 3.96 $\mu$s
\cite{nvidia_nvqlink_blog} \\
Transfer model & $T_{\text{link}}(B)=3.84\,\mu\mathrm{s}
+ B/(50\,\mathrm{GB/s})$ before chunking. Chunked transfers charge latency per
chunk when endpoint buffers are smaller than the payload. \\
\midrule
\multicolumn{2}{@{}l}{\textbf{Quantum annealer (QA)}}\\
Programming time $T_{\text{program}}$ & 17.7 ms
\cite{dwave_qpu_properties} \\
Anneal time $T_{\text{anneal}}$ & 0.5--2000 $\mu$s
\cite{dwave_qpu_properties} \\
Readout time $T_{\text{readout}}$ & 17--265 $\mu$s
\cite{dwave-annealing} \\
Per-sample delay $T_{\text{delay}}$ & 20.6 $\mu$s
\cite{dwave_qpu_properties} \\
Reads per QA job & workload-dependent, 200--2200 in the workload table. \\
QA job-time structure &
$T_{\text{program}}+N_{\text{reads}}
(T_{\text{anneal}}+T_{\text{readout}}+T_{\text{delay}})$, refined by
backend-specific QPU-access fits in the hardware-grounded profiles. \\
\midrule
\multicolumn{2}{@{}l}{\textbf{Digital quantum computer (DQC)}}\\
Sub-job overhead $T_{\text{overhead}}$ & 2.0 s
\cite{ibm_estimate_runtime} \\
Repetition delay & 250 $\mu$s \cite{ibm_estimate_runtime} \\
Gate-time prior & 100 ns \cite{ibm-quantum-qpu-information} \\
Circuit duration prior & $T_{\text{gate}}\times\text{circuit width/depth
proxy}$, replaced in the calibrated model by {\leavevmode circuit family (circuits with a common structure)}, depth,
transpiled depth, gate counts, qubits, and shots. \\
Circuits per DQC job & workload-dependent, 8--128 in the original HCU
workload ranges and refined by the current workload registry. \\
Shots per circuit & workload-dependent, 50--1200 in the original HCU ranges
and refined by current generated circuits. \\
DQC job-time structure &
$T_{\text{overhead}}+(T_{\text{circuit}}+\text{rep\_delay})\cdot N_{\text{exec}}$,
with IBM-reported quantum seconds used as the calibrated service target. {\leavevmode Here, $T_{\text{circuit}}$ is circuit duration, $\text{rep\_delay}$ is repetition delay, and $N_{\text{exec}}$ counts circuit executions, including repeated shots.} \\
\midrule
\multicolumn{2}{@{}l}{\textbf{Data movement and constraints}}\\
QA input/output payloads & 4--40 KB / 4--32 KB in the original compact
parameterization. Current workload table exposes per-workload payload ranges. \\
DQC input/output payloads & 4--200 KB / 4--40 KB in the original compact
parameterization. Current workloads include larger generated payloads such as
portfolio and PES scan configurations. \\
QA$\rightarrow$DQC direct path & not available by default. Cross-modal
workflows are host/controller mediated unless a topology explicitly declares a
controller cross-link or direct QPU link. \\
QPU lock semantics & exclusive during physical quantum execution and
programming/readout intervals charged by the backend profile. \\
\bottomrule
\end{tabular}
\end{table*}

\subsection{Controllers and Links as First-Class Resources}

Controllers are explicit resources in the graph, not hidden constants inside a
QPU service time. A QA controller can serialize embedding submission,
programming, readout, and result return. A DQC controller can serialize circuit
submission, batched execution setup, readout, and feedback. Links are likewise
explicit edges with latency, bandwidth, and routing constraints. Treating these
components as first-class is what lets HybridQC~distinguish a QPU bottleneck from a
controller or interconnect bottleneck.

This distinction is essential for integrated HCU studies. If a workload adds
DQC replicas but keeps one DQC controller, the simulator can expose the
controller as the next limiting resource. If a workload grows output payloads,
the simulator can charge the relevant link rather than hiding the transfer in a
generic postprocessing constant. The evaluation uses this capability directly:
the topology pair sweeps in Section~\ref{sec:results} show that QA-control
replication and DQC-control replication optimize different objectives.

\subsection{Configuration Interface}

{\leavevmode The configuration interface makes architectural choices explicit. A user can add or remove resources, change controller concurrency, alter link bandwidth, select backend profiles, and switch scheduling policies without rewriting the simulator. Table~\ref{tab:hcu_parameters_exact_app} records the baseline timing fields and units. Input examples and execution instructions are in the project documentation~\cite{hybridqc_project}.}

\subsection{Capacity, Transfer, and Feasibility Semantics}

Transfers are topology-aware. If a payload of $B$ bytes crosses an edge with
bandwidth $\beta$ {\leavevmode bytes/s} and latency $\ell$ {\leavevmode seconds}, the nominal transfer time is
$T_{\mathrm{edge}}(B)=\ell+B/\beta$. Large-payload experiments cannot
silently pass oversized data as one event. If the endpoint memory or
controller buffer admits at most $C$ bytes per transfer chunk, HybridQC~charges
latency per chunk:
\begin{equation}
T_{\mathrm{edge}}(B)=\left\lceil B/C \right\rceil \ell + B/\beta.
\end{equation}
If a memory requirement exceeds capacity but the stage is streamable, the
simulator models repeated chunks and sustained capacity use. If a quantum
footprint exceeds the target backend and no decomposition is enabled, the configuration
is marked infeasible rather than converted into a misleading runtime.
{\leavevmode The supporting workload-scaling results~\cite{hybridqc_supplementary_data} use} this distinction when reporting
workload-scaling configurations that complete, saturate, or become infeasible. These
semantics are part of the HCU abstraction rather than an evaluation afterthought:
without them, workload-scaling experiments would silently undercharge transfer and
memory pressure.

\subsection{{\leavevmode Cross-Modal Handoff and Service Decomposition}}
Workloads are expressed as staged hybrid workflows rather than single quantum
calls. The canonical modes are QA-only optimization, DQC-only circuit
execution, QA-to-DQC refinement, and QA with classical mediation. {\leavevmode Cross-modality jobs combine QA and DQC and are host- and controller-mediated by default}: QA results return through
readout, controller, host/GPU, and transfer stages before DQC refinement is
submitted. This prevents the simulator from assuming a direct QA-to-DQC path
unless the topology explicitly provides one.

Controllers are finite-capacity service resources. They can model host relay,
controller-side processing, controller-to-controller cross-links, cryogenic
edge control, hierarchical controller meshes, QPU-direct paths, and full
bypass modes. A quantum segment is therefore decomposed into transfer-in,
controller service, physical execution, controller readout/return, and
transfer-out terms:
\begin{align}
{\leavevmode T_{\mathrm{QA,segment}}} &=
T_{\mathrm{in,QA}} + T_{\mathrm{QA,ctrl}} + T_{\mathrm{QA,exec}}
  \notag\\
&\quad + T_{\mathrm{QA,ctrl,out}} + T_{\mathrm{out,QA}}, \\
{\leavevmode T_{\mathrm{DQC,segment}}} &=
T_{\mathrm{in,DQC}} + T_{\mathrm{DQC,ctrl}} + T_{\mathrm{DQC,exec}}
  \notag\\
&\quad + T_{\mathrm{DQC,ctrl,out}} + T_{\mathrm{out,DQC}} .
\end{align}
{\leavevmode Here $T_{\mathrm{in}}$ and $T_{\mathrm{out}}$ are transfer times, $T_{\mathrm{ctrl}}$ is controller service, $T_{\mathrm{exec}}$ is physical execution, and $T_{\mathrm{ctrl,out}}$ is readout/return service, all in seconds. The sums give segment service time. Resource contention adds waiting, and independent segments can overlap. Quantum error correction (QEC) adds repeated error-detection measurements, classical decoding, and corrective feedback. These protection stages are disabled in the evaluation, which studies application, control, transfer, and classical-processing costs.}

\paragraph{{\leavevmode Topology and profile interaction.}}
The link model supports host-side links such as NVQLink, PCIe Gen4/Gen5,
InfiniBand NDR400, and 100G Ethernet, plus controller/QPU links such as
cryogenic coax, cryogenic CMOS, and photonic-control profiles. Cross-modality
connectivity can be mediated-only, controller-cross-linked, or direct if the
user declares such a path. Resource replication creates additional resources
with the same profile and connects replicas according to the declared routing
mode.

{\leavevmode A backend profile supplies the service time of a reservation; the resource graph determines the device, controller, and links available to that stage. Changing a route changes its transfer and controller costs without changing the workload dependencies. The grounded QA and DQC devices are reserved exclusively during the physical occupancy interval: the service measurements represent a submitted device job, so overlapping independent jobs on the same device would require a different sharing model.}

\section{HybridQC: Workloads, Grounding, and Simulator Design}
\label{sec:workloads}

\subsection{System-Level Multi-Stage Simulation}
HybridQC~does not represent a hybrid workload as one quantum-service number because
that would erase the resources that make topology matter. Instead, each job is
lowered into a system-level multi-stage DAG whose nodes carry a resource type,
runtime model, memory requirement, payload size, routing metadata, and
predecessor dependencies. Classical stages reserve CPU/GPU nodes, QA and DQC
stages reserve backend-profiled quantum nodes and controllers, and transfer
stages move data across explicit links in the HCU topology. Problem mapping,
embedding/transpilation, controller programming, readout, transfers,
reconstruction, and postprocessing are explicit stages rather than hidden
constants. {\leavevmode Compiler, circuit-cutting, and QEM costs can be attached as parameterized stages. Circuit cutting divides a circuit into smaller circuits and adds classical work to reconstruct the requested result. QEM stages charge the extra measurements, transfers, or classical processing specified by the workload.} The
evaluation below does not claim a particular compiler, cutting algorithm, or
mitigation implementation. In the implementation, each workload DAG is emitted
as typed stage records that preserve resource needs, payload sizes,
dependencies, and runtime-profile hooks.

This is why two workloads with similar QPU
service time can have different end-to-end behavior. A portfolio workflow may
be QA-to-DQC and transfer-heavy, a VQE workflow may be DQC-iteration-heavy, and
a bin-packing workflow may mainly stress QA sampling and QA-controller staging.
{\leavevmode Table~\ref{tab:workload_parameter_ranges} and the scaling definitions below provide the inputs needed to interpret the workload-scaling experiments.}

\begin{table}[t]
\centering
\caption{Workload families lowered by HybridQC. Representative workloads are
grounded in hybrid QA/DQC optimization and chemistry workloads. {\leavevmode Table~\ref{tab:workload_parameter_ranges} gives the quantitative inputs, and the scaling definitions below specify how they grow.}}
\label{tab:workload_families}
\Description{Rows group workloads into annealing-heavy, digital-quantum-heavy, cross-modality, and mixed bundles. Columns name representative application families and the stages in their execution graphs.}
\setlength{\tabcolsep}{3.5pt}
\renewcommand{\arraystretch}{1.12}
\small
\begin{tabular}{@{}p{0.18\linewidth}p{0.28\linewidth}p{0.42\linewidth}@{}}
\toprule
\textbf{Class} & \textbf{Families} & \textbf{Representative stage graph} \\
\midrule
QA-heavy & bin packing~\cite{de_Andoin_2022}, vehicle
routing~\cite{maciejunes2025solvinglargescalevehiclerouting} &
preprocessing, embedding, programming, QA sampling, result transfer \\
\addlinespace[0.2em]
DQC-heavy & VQE~\cite{peruzzo2014vqe}, PES
scan~\cite{Brown_2024}, job
shop~\cite{sawamura2025quantumclassicalhybridalgorithmusing}, feature
selection & transpilation, circuit execution, readout, reconstruction \\
\addlinespace[0.2em]
Cross-modality & portfolio~\cite{Morapakula_2025}, unit commitment,
transmission switching~\cite{ellinas2024hybridquantumclassicalalgorithmmixedinteger},
pipeline/SQC~\cite{pipeline_optimization} & QA warm-start or search,
transfer, DQC refinement, classical postprocessing \\
\addlinespace[0.2em]
Mixed bundles & combinations of the above & independent typed DAGs
co-scheduled on the same HCU \\
\bottomrule
\end{tabular}
\end{table}

\begin{figure}[t]
  \centering
  \sbox{\hybridqcfigurebox}{\includegraphics{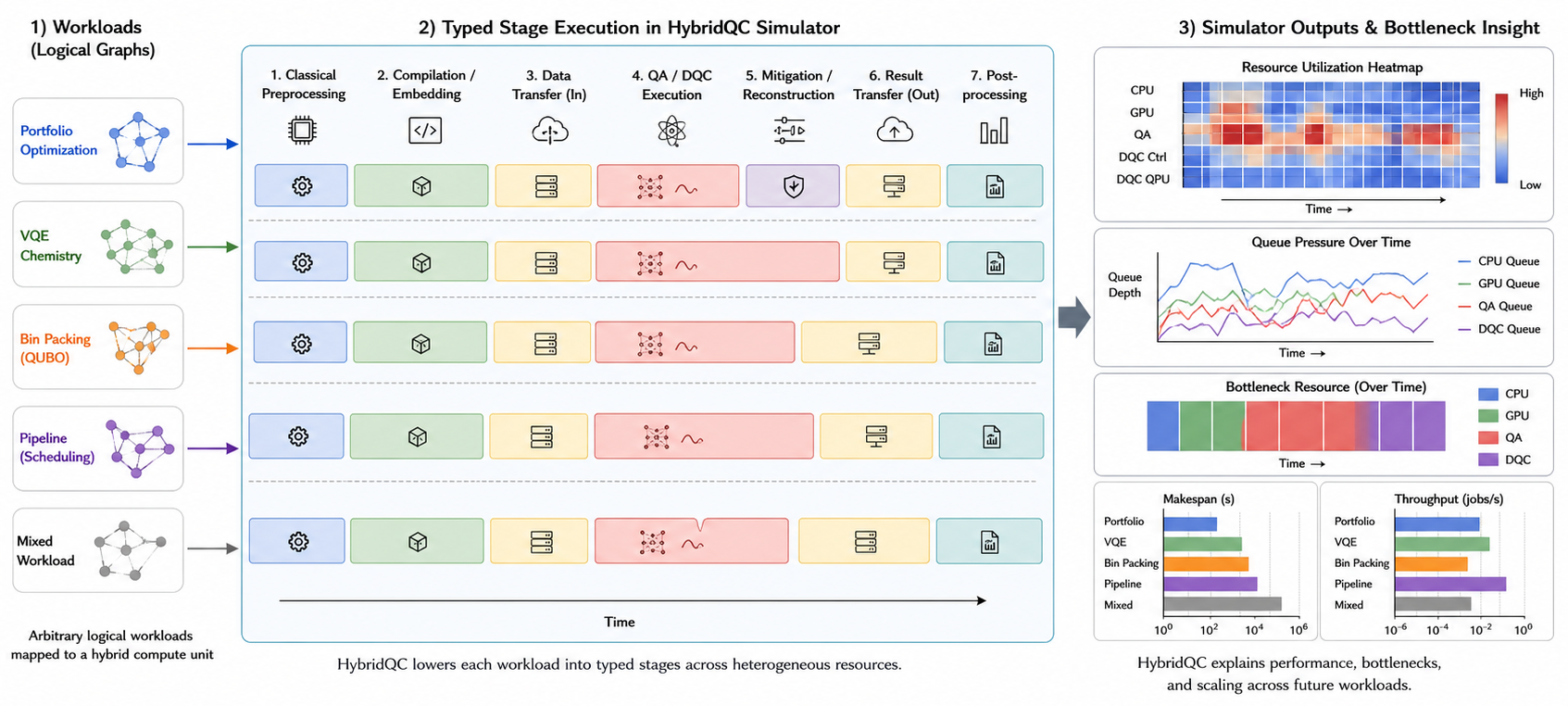}}%
  \includegraphics[trim={0.695\wd\hybridqcfigurebox} 0pt 0pt {0.06\ht\hybridqcfigurebox},clip,width=\columnwidth]{paper_sections/figures/workload_stage_bottleneck_ai_cropped.png}
  \caption{{\leavevmode System-level outputs from staged workload execution. The panels illustrate resource utilization, queue pressure, bottleneck movement, makespan, and throughput. Each quantity is collected while the typed stages contend for the resources and links in the configured HCU.}}
  \label{fig:workload_stage_bottleneck}
  \Description{An illustrative simulator-output dashboard contains a resource-utilization heatmap, queue-depth curves over time, a timeline of the resource acting as the bottleneck, and horizontal bar charts of makespan and throughput by workload. The heatmap and queue curves distinguish CPU, GPU, quantum-annealing, and digital-quantum resources. Together the panels show how the simulator connects changing resource activity and queue pressure to application-level outcomes.}
\end{figure}

{\leavevmode Figure~\ref{fig:workload_stage_bottleneck} summarizes the metrics collected during staged execution.} The same logical job can
generate CPU/GPU preprocessing, problem-mapping, quantum service, controller
work, transfers, and reconstruction stages, and each of those stages can wait
on a different resource.
\subsection{Workload Parameters}
The second design choice is to keep workload parameters visible rather than
collapsing them into one service-time multiplier. Table~\ref{tab:workload_families}
groups the workload suite by architectural stress so that each workload family maps to a
different region of the HCU design space. QA-heavy workloads primarily exercise
embedding, programming, sampling, and QA result movement. DQC-heavy workloads
stress transpilation, circuit execution, readout, and reconstruction.
Cross-modality workloads are the most important for HCU studies because they
use both quantum modalities and move intermediate data between them. Mixed
bundles then co-schedule these workloads on the same topology, exposing
contention that runs of one workload family can hide.

The quantitative ranges in {\leavevmode Table~\ref{tab:workload_parameter_ranges}}
make the workload-scaling experiments interpretable. Shots, circuit count,
depth, logical qubits, and payload size are not interchangeable knobs: QA shots
primarily increase annealing service and readout demand. DQC circuit count and
shots increase quantum-service demand. {\leavevmode Depth, the number of sequential gate layers, changes circuit duration}, and
input/output payloads stress links, buffers, and memory. {\leavevmode The compute and payload fields in Table~\ref{tab:workload_parameter_ranges} specify the inputs to these stages.}

\begin{table*}[t]
\centering
\caption{Quantitative workload parameters used by the simulator. We sample each
range uniformly per job instance. Columns group \emph{compute intensity}
(QA/DQC execution parameters) and \emph{communication intensity} (payload sizes
on QA/DQC edges). {\leavevmode ``params'' denotes the circuit parameter count; ``len.'' is the DQC circuit-duration input in microseconds.} Output payloads scale QA/DQC result bytes by $16\times$ to
represent multiple measurement data that need to be transferred to the
classical host. The workload families follow QA optimization, VQE/QAOA,
PES, scheduling, routing, power-system, portfolio, and pipeline optimization
workloads~\cite{de_Andoin_2022,maciejunes2025solvinglargescalevehiclerouting,peruzzo2014vqe,farhi2014qaoa,Brown_2024,sawamura2025quantumclassicalhybridalgorithmusing,ellinas2024hybridquantumclassicalalgorithmmixedinteger,Morapakula_2025,pipeline_optimization}.}
\label{tab:workload_parameter_ranges}
\Description{Eleven workload rows list quantum execution inputs and communication payloads. Columns report annealing shots, circuit count, digital-quantum shots, qubits, depth, parameter count, circuit duration, and input and output bytes for each quantum modality.}
\setlength{\tabcolsep}{3.5pt}
\renewcommand{\arraystretch}{1.05}
\scriptsize
\begin{lrbox}{\hybridqctablebox}
\begin{tabular}{l r r r r r r r r r r r}
\toprule
\multicolumn{1}{c}{} &
\multicolumn{7}{c}{\textbf{Compute / quantum-work parameters}} &
\multicolumn{4}{c}{\textbf{Communication payloads (bytes)}} \\
\cmidrule(lr){2-8}\cmidrule(lr){9-12}
\textbf{Workload} &
\textbf{QA shots} &
\textbf{\#circuits} &
\textbf{DQC shots} &
\textbf{qubits} &
\textbf{depth} &
\textbf{params} &
\textbf{len. ($\mu$s)} &
\textbf{QA in} &
\textbf{QA out} &
\textbf{DQC in} &
\textbf{DQC out} \\
\midrule
Unit Commitment
& 1200--2000 & 32--64 & 400--800 & 36--60 & 1000--1600 & 80--200 & 100--160
& 24k--40k & 256k--448k & 80k--140k & 320k--576k \\
Transmission Switching
& 700--1200 & 64--128 & 600--1200 & 40--72 & 1200--2000 & 100--220 & 120--200
& 20k--36k & 256k--448k & 120k--200k & 384k--640k \\
Cloud Scheduling
& 500--900 & 48--96 & 300--700 & 32--56 & 800--1400 & 70--180 & 80--140
& 12k--24k & 192k--384k & 90k--160k & 288k--480k \\
Pipeline (Streaming)
& 1000--1600 & 48--96 & 600--1200 & 36--64 & 1000--1800 & 90--220 & 100--180
& 20k--40k & 256k--512k & 100k--180k & 384k--640k \\
Portfolio Optimization
& 200--1200 & 16--64 & 150--600 & 20--48 & 500--1600 & 40--120 & 50--160
& 8k & 256k & 64M & 256M \\
Job Shop Scheduling
& -- & 64--128 & 400--900 & 28--52 & 900--1600 & 60--140 & 90--160
& -- & -- & 60k--120k & 256k--448k \\
Feature Selection
& -- & 32--72 & 500--1000 & 24--48 & 700--1200 & 50--140 & 70--120
& -- & -- & 50k--90k & 256k--448k \\
Bin Packing
& 600--1000 & -- & -- & -- & -- & -- & --
& 8k--16k & 128k--256k & -- & -- \\
Vehicle Routing
& 1400--2200 & -- & -- & -- & -- & -- & --
& 16k--32k & 192k--384k & -- & -- \\
VQE
& 128--384 & 25--49 & 96--256 & 12--32 & 300--1200 & 12--24 & 40--100
& 8k--24k & 128k--384k & 64k--256k & 256k--1024k \\
PES scan
& 0 & 1 & 512 & 20 & 100 & 200--400 & 0
& 0 & 0 & 200M & 800M \\
\bottomrule
\end{tabular}
\end{lrbox}%
\ifdim\wd\hybridqctablebox>\textwidth
  \resizebox{\textwidth}{!}{\usebox{\hybridqctablebox}}%
\else
  \usebox{\hybridqctablebox}%
\fi
\end{table*}

\subsection{Scaling Semantics}

The third design choice is to scale workloads along explicit axes: data,
shots, depth, circuit count, circuit-shot product, circuit-shot-depth product,
and all fields. The workload-wide scale factor multiplies the workload's
compute and payload parameters before simulation, while communication-only
scaling can separately stress input and output movement. This wording matters:
100$\times$ data-only growth is not equivalent to 100$\times$ growth in
circuits, shots, and depth. Keeping these axes separate lets the simulator ask
which hardware path becomes limiting as workloads grow. When large
payloads no longer fit in one transfer or controller buffer, the chunked
transfer semantics from Section~\ref{sec:model} apply.
{\leavevmode The definitions below specify these scale modes; supporting data~\cite{hybridqc_supplementary_data} retains the full completed and unfinished 100$\times$ results.}

The evaluation distinguishes:
\begin{itemize}[leftmargin=*, itemsep=0.1em]
  \item data-only scaling, which increases payload and memory pressure.
  \item shot-only scaling, which increases repeated DQC/QA measurement demand.
  \item depth-only scaling, which increases circuit service time.
  \item circuit-only scaling, which increases the number of submitted quantum
  kernels.
  \item circuit-shot and circuit-shot-depth scaling, which multiply several
  DQC service axes together.
  \item all-field scaling, which is allowed to become infeasible when device or
  memory capacities are exceeded.
\end{itemize}

\subsection{Hardware-Grounded Runtime Profiles}
\label{sec:grounding}

\subsubsection{Backend-Specific Profiles}

HybridQC~uses backend-specific runtime profiles because the same stage graph should
execute differently on different machines. Generic QA and DQC profiles provide
defaults. Backend-specific profiles refine those defaults with measured data.
{\leavevmode The measured quantum computers are D-Wave Advantage 1 and Advantage 2 annealers and IBM Kingston, Marrakesh, and Fez gate-based devices~\cite{dwave_qpu_properties,ibm_quantum_computer_catalog}.} Backend objects store not only fitted
coefficients but also hardware metadata such as qubit counts, timing fields,
embedding status, transpiled circuit properties, and provider-result fields.
This makes profiles auditable and re-groundable as new measurements arrive.

{\leavevmode Complete hardware measurements, fitted parameters, and results are available in the supporting data~\cite{hybridqc_supplementary_data}.}

\subsubsection{Physical Occupancy versus Cloud Wall Time}
For D-Wave, the physical occupancy target is QPU access time. Programming,
anneal, readout, delay, sampling, embedding status, and wall time are retained
as separate fields. For IBM, the physical occupancy target is IBM-reported
quantum seconds, fitted from circuit count, shots, qubits, depth, transpiled
depth, two-qubit gates, backend, and circuit family. Queue and provider wall
fields are used only in optional cloud-proxy mode. Thus the same hardware
records can validate today's cloud-access behavior without forcing future
integrated HCUs to inherit cloud overhead as a device constant. {\leavevmode The equations below define the main timing features; the supporting data~\cite{hybridqc_supplementary_data} retains the full target/predictor mapping.}

\subsubsection{Grounding Campaigns and Simulation Interface}
The grounding campaigns are designed around the simulator interface rather than
around isolated benchmark curves. Broad calibration grids span problem size,
circuit shape, backend, shot/read count, and timing fields so that backend
profiles have enough variation to fit service-time coefficients. {\leavevmode Measurements for portfolio, pipeline, PES-scan, and VQE workloads are retained separately to check fitted profiles on application-derived inputs.}
After fitting, each backend profile supplies service-time estimates to the
typed stages described in Section~\ref{sec:workloads}: stage graphs determine
\emph{where} a job needs service, while backend profiles determine
\emph{how long} each QA or DQC reservation occupies the HCU. Integrated-mode
runs use calibrated physical service. Cloud-proxy runs add provider overhead
when reproducing today's cloud execution.

\paragraph{D-Wave QA physical-access model.}
{\leavevmode The baseline QA physical-access time $T_{\mathrm{QA,access}}$ is}
\begin{equation}
{\leavevmode T_{\mathrm{QA,access}}} =
T_{\mathrm{program}}+
N_{\mathrm{reads}}\left(T_{\mathrm{anneal}}+T_{\mathrm{readout}}+T_{\mathrm{delay}}\right),
\end{equation}
{\leavevmode where $N_{\mathrm{reads}}$ is the sample count. $T_{\mathrm{program}}$ is the once-per-job programming time, and $T_{\mathrm{anneal}}$, $T_{\mathrm{readout}}$, and $T_{\mathrm{delay}}$ are the per-read annealing, readout, and delay times, in seconds.} Backend-specific profiles refine these
terms with measured D-Wave QPU-access time. Provider wall time is retained as
a separate cloud-overhead observable and is not treated as physical QPU
occupancy.

\paragraph{IBM Runtime quantum-seconds model.}
For IBM backends, the fitted target is IBM-reported quantum seconds rather than
client wall time. The backend coefficient table in the supporting data~\cite{hybridqc_supplementary_data} uses the
feature form
\begingroup
\begin{equation}
\begin{split}
\widehat T_{\mathrm{IBM}} &= \alpha_0+\alpha_C N_C+\alpha_S\frac{S}{10^3}\\
&\quad+\alpha_D\frac{S\sum_{i=1}^{N_C}D_i}{10^6}
+\alpha_{QD}\frac{S\sum_{i=1}^{N_C}Q_iD_i}{10^8},
\end{split}
\end{equation}
where $\widehat T_{\mathrm{IBM}}$ predicts quantum seconds for $N_C$ circuits with $S$ shots per circuit. Circuit $i$ has depth $D_i$ and qubit count $Q_i$. The backend-specific coefficients weight the intercept ($\alpha_0$), circuit count ($\alpha_C$), shots ($\alpha_S$), depth ($\alpha_D$), and qubit-depth product ($\alpha_{QD}$). Each term contributes seconds. The denominators scale the predictors.
\endgroup

{\leavevmode Table~\ref{tab:validation_error} reports the fit to D-Wave Advantage 2 training measurements and the validation on D-Wave Advantage 1 workload measurements. Each IBM result evaluates the fit on measurements from that backend. The table gives the measurement counts.}

\subsection{Simulator Design and Policy Co-Design}
\label{sec:sim_design}

\subsubsection{Execution Pipeline}

The execution engine is designed to preserve the same resource semantics
introduced above. HybridQC~takes a workload list, applies workload-scaling
parameters, lowers each job into typed stages, routes payloads through the
HCU~graph, and runs a discrete-event simulation until all jobs complete or the
horizon is reached. Stage starts and completions update resource state, queues,
memory occupancy, and metric accumulators. This design makes end-to-end
workload behavior visible: a DQC-heavy job can still be delayed by
problem-preparation, data movement, controller availability, or reconstruction.

The implementation has two cooperating layers. The Python pipeline is the
canonical model: it owns workload construction, backend-profile loading,
subroutine expansion, measurement records, and experiment orchestration. The
Rust native engine executes the typed event path used for large sweeps. This
split keeps the model inspectable while allowing the hot event loop, typed
queues, and metric accumulation to run without Python object churn.
Section~\ref{sec:evaluation} defines the implementation-aligned diagnostic
metrics used in the evaluation tables.

\paragraph{{\leavevmode Readiness, reservation, and release.}}
{\leavevmode A stage becomes ready when its declared predecessor conditions are satisfied. It can start when the required resource capacities and memory are available; the scheduler chooses among stages that can run. A stage can require several resources, such as a QPU and its controller or a transfer route and endpoint buffers. Their reservations prevent conflicting work from using the same capacity during that service interval. Completion releases the stage's reservations and updates successor readiness. Independent stages can overlap when their resource requirements fit simultaneously. Scheduling and state-collection overhead are assumed negligible in the evaluated simulations; stage service and queueing determine simulated elapsed time.}

\paragraph{{\leavevmode QA-to-DQC overlap example.}}
{\leavevmode Consider a QA producer that supplies two result batches to a DQC refinement stage through the host and controllers. With batch-level dependencies, transfer and preparation of the first batch can begin after that batch is available, while the producer continues its remaining QA work. DQC refinement starts after the first batch's required preparation and transfer stages finish and the DQC device and controller are available. The second batch follows the same path. A full-producer dependency instead delays the consumer until both batches are ready. Thus overlap follows the declared dependency granularity and available resources: a shared busy controller or link can still serialize the handoff. This is an illustrative execution example, not an additional experiment.}

\subsubsection{Typed Stages and Subroutines}

Problem preparation, embedding/transpilation, QA execution, DQC execution,
controller programming/readout, transfers, reconstruction, and postprocessing
are represented as typed stages. This is a design choice because these
subroutines consume different resources and can become bottlenecks
independently. The typed representation avoids treating subroutines as opaque
constants and makes the implementation testable: backend profiles can be
validated against QA/DQC timing data, stage expansions can be checked
independently, and policies can be tested against explicit queue states.
Compiler-specific, circuit-cutting, and QEM methods are extensibility points
that can be modeled as additional parameterized stages, but they are not
method-specific claims in this evaluation.

\subsubsection{Scheduling as a Topology Knob}
\label{sec:policies}

Topology and scheduling are coupled: the topology defines which resources can
become scarce, while the policy decides which ready stage receives a scarce
controller, QPU, link, or classical resource next. HybridQC~therefore treats
policy choice as part of the HCU configuration rather than as an implementation
detail.

The evaluated policies span different scheduling aims. FIFO is the arrival-order baseline. Timeout batching releases QA and DQC
controller batches once either a size or wait threshold is reached.
Hybrid Overhead-Balanced Adaptive (HOBA) uses queue-aware dispatch and batching to
protect constrained quantum paths. Adaptive Batch-Optimal (ABO) uses
arrival-rate-aware batching, while Greedy Max Pressure (GMP) and
Max-Pressure {\leavevmode rank work using resource-demand differences}. 
{\leavevmode Table~\ref{tab:policy_state_app} and the scoring definitions below give the policy-visible state and dispatch behavior. Supporting data~\cite{hybridqc_supplementary_data} retains the complete mixed-policy results.}

{\leavevmode Let $q$ be an eligible stage, $r(q)$ its resource, and $r_{\mathrm{next}}(q)$ its next resource. Predicted service time $s(q)$, queue age $a(q)$, and estimated outstanding service $L_r$ use seconds. Dispatch respects resource availability and dependencies. FIFO orders stages by arrival time. The pressure dispatch rules are given below, while ABO minimizes the batch-cost objective.}

\begin{table*}[t]
  \centering
  \footnotesize
  \begin{lrbox}{\hybridqctablebox}
  \begin{tabular}{@{}p{0.17\textwidth}p{0.20\textwidth}p{0.24\textwidth}p{0.17\textwidth}p{0.16\textwidth}@{}}
    \toprule
    Policy & State used & Dispatch rule & Best use case & Failure mode \\
    \midrule
    FIFO & enqueue order & oldest ready stage first &
    deterministic baseline and low-contention queues & ignores bottleneck and
    controller pressure \\
    Timeout Batching & age, queue length, batch limit & release a batch when
    size or timeout threshold is met & controller setup amortization & can add
    unnecessary waiting under light load \\
    Adaptive Batch-Optimal (ABO) & arrival-rate estimate, setup cost, batch cap
    & choose QA/DQC batch sizes from an overhead/wait tradeoff &
    controller setup amortization & can over-batch when arrival estimates are
    stale \\
    Greedy Max Pressure (GMP) & {\leavevmode resource-demand difference, age, downstream backlog,
    feed urgency, service time} & {\leavevmode first eligible on QPU/controller queues and formed batches, otherwise maximize $S_{\mathrm{GMP}}$} & mixed workloads with persistent downstream queues & can
    misweight resources if pressure is noisy \\
    Max-Pressure & {\leavevmode local resource-demand difference and age} & {\leavevmode maximize $S_{\mathrm{MP}}$ for ordinary stages and $L_{r(q)}+0.01a(q)$ for formed batches} & high-contention pipelines & can starve low-pressure but
    important work \\
    Hybrid Overhead-Balanced Adaptive (HOBA) & queue state, batch caps,
    feed-QA/feed-DQC urgency & combine queue-aware dispatch with QA/DQC
    batching & controller-mediated hybrid queues & depends on signal quality
    and tuning \\
    \bottomrule
  \end{tabular}
  \end{lrbox}%
\ifdim\wd\hybridqctablebox>\textwidth
  \resizebox{\textwidth}{!}{\usebox{\hybridqctablebox}}%
\else
  \usebox{\hybridqctablebox}%
\fi
  \caption{Policy state and dispatch interpretation. These policies are HCU
  configuration knobs. The main contribution is evaluating policy and topology
  together under explicit controller and link contention.}
  \label{tab:policy_state_app}
\Description{Six scheduling policies are compared by the state they use, dispatch rule, intended use case, and failure mode. The rows cover FIFO, timeout batching, adaptive batching, greedy max pressure, max pressure, and hybrid overhead-balanced adaptive scheduling.}
\end{table*}

\paragraph{Greedy Max Pressure (GMP) and Max-Pressure.}
\begingroup
Define $\Delta L(q)=L_{r(q)}-L_{r_{\mathrm{next}}(q)}$, with zero successor demand for terminal stages. The ranking scores are
\begin{equation}
\begin{split}
S_{\mathrm{MP}}(q)&=\Delta L(q)+0.01a(q),\\
S_{\mathrm{GMP}}(q)&=\Delta L(q)+\eta a(q)-\kappa s(q)+F(q).
\end{split}
\end{equation}
The dimensionless weights $\eta$ and $\kappa$ control aging and service-time preference. $F(q)$ is the configured bonus for feeding an underfilled QA/DQC path, on the same seconds-based score scale.

Table~\ref{tab:policy_state_app} gives the dispatch rules for ordinary stages and formed batches.
\endgroup

\paragraph{Adaptive Batch-Optimal (ABO).}
ABO is the arrival-rate-aware batching policy. For a controller setup overhead
$O$ {\leavevmode in seconds}, estimated arrival rate $\lambda$ {\leavevmode in requests/s}, {\leavevmode dimensionless} waiting weight $W$, and candidate batch
size {\leavevmode $b$ requests}, the simulator selects the feasible batch size that minimizes
\begin{equation}
{\leavevmode C_{\mathrm{batch}}(b)}=\frac{O}{b}+W\frac{b-1}{2\lambda},
\end{equation}
subject to the QA/DQC batch caps and hard timeout settings. {\leavevmode The first term is setup overhead per request. The second estimates the waiting cost of collecting a batch.} The risk is
over-batching when the arrival-rate estimate becomes stale relative to the
current workload mix.

All policies use deterministic tie-breaking in the simulator so that repeated
runs are comparable. The scoring intuition and equations explain the policy effects reported below.

\section{Evaluation and Results}
\label{sec:evaluation}
\label{sec:results}

\subsection{Methodology and HybridQC{} Design Questions}

The evaluation separates \emph{validation} from \emph{forecasting}. Validation
asks whether the multi-stage runtime models used by HybridQC~are compatible with
live hardware measurements. Forecasting asks what those grounded models imply
for full hybrid workloads running on proposed tightly integrated topologies.

The evaluated workload stages include problem preparation, embedding or
transpilation, controller programming/readout, QA/DQC execution, transfers,
reconstruction, postprocessing, memory capacity, queues, and scheduling
policies. Compiler-specific, circuit-cutting, and QEM algorithms are supported
as parameterized stage extensions, but are not evaluated as named methods in
this paper.
The workload suite is summarized in Table~\ref{tab:workload_families}:
QA-heavy optimization, DQC-heavy circuit workloads, cross-modality QA-to-DQC
workflows, and mixed bundles. {\leavevmode Table~\ref{tab:workload_parameter_ranges}} gives the full numeric parameter
ranges.

We organize the evaluation around the following questions.
\begin{itemize}[leftmargin=*, itemsep=0.15em, topsep=0.2em]
  \item \textbf{V1: Hardware grounding.} Can backend-specific QA and DQC
  service models fit measured D-Wave access time and IBM-reported quantum
  seconds well enough to support system-level and multi-stage simulation?
  \item \textbf{E1: Topology balance.} When additional hardware is available,
  does adding one resource class suffice, or do workloads require balanced
  scaling across classical, quantum, controller, and link resources?
  \item \textbf{E2: Scheduling.} Do queueing policies materially affect
  throughput and bottleneck migration for mixed hybrid workloads?
  \item \textbf{S1: Workload scaling.} Which scaling axes are benign and which
  axes produce architectural saturation at 10$\times$ and 100$\times$ growth?
\end{itemize}

{\leavevmode Supporting data~\cite{hybridqc_supplementary_data} preserves the deployment-mode, mixed-workload, load-amplification, and sensitivity results.}

\subsection{Implementation and Experiment Inputs}
The experiments use the Python configuration pipeline and Rust execution engine described in Section~\ref{sec:sim_design}. IBM grounding uses Qiskit/Runtime
artifacts~\cite{qiskit_zenodo,ibm_estimate_runtime,ibm_sessions}; D-Wave uses
Ocean SampleSets~\cite{dwave_ocean_sdk}. Each experiment specifies a workload bundle,
backend/profile metadata, scaling or shot/read parameters, HCU graph, policy,
and timing fields. The same workload families generate the IBM circuits and
D-Wave {\leavevmode quadratic unconstrained binary optimization (QUBO) problems} used for grounding, keeping calibration tied to simulated
workloads. {\leavevmode Section~\ref{sec:policies} summarizes policy state and scoring objectives. Hardware measurements, backend metadata, and additional results are available in the supporting data~\cite{hybridqc_supplementary_data}. The configuration listing is in the project documentation~\cite{hybridqc_project}.}

{\leavevmode For the single-resource replica sweep, each of the 11 workload families runs separately with 30 jobs at 1$\times$ workload scale under the HOBA policy. Compilation, cutting, and QEM stages are enabled. Each comparison varies one resource count while keeping the other counts and link profiles fixed. The grounding campaigns and occupancy targets are described in Section~\ref{sec:grounding}.}

For each backend profile, we report MAPE and
coefficient of determination ($R^2$) on the physical-occupancy target used by
the simulator. For measured target $y_i$ (hardware runtime) and prediction
$\hat{y}_i$ (HybridQC~runtime),
$\mathrm{MAPE}=(100/n)\sum_i |(y_i-\hat{y}_i)/y_i|$, computed on positive
timing targets, and
$R^2=1-\sum_i(y_i-\hat{y}_i)^2/\sum_i(y_i-\bar{y})^2$, which measures the
fraction of target variance explained by the fitted profile. {\leavevmode Here $i$ indexes the $n$ measurements, and $\bar y$ is their mean measured target.} These definitions
are used by the grounding anchors below.

\subsection{Evaluation Metrics}
HybridQC~reports only implementation metrics used in the results. Let $J$ be the
submitted {\leavevmode job set}, $J_c$ the {\leavevmode subset of} completed jobs, $A_j$ and $C_j$ the arrival and
completion times {\leavevmode of job $j$}, and {\leavevmode $T_0=\min_{j\in J} A_j$}. The reported run makespan is
\texttt{total\_time\_s}: $T_{\mathrm{run}}=\max_{j\in J_c}C_j-T_0$. Per-job
completion time is $T_j=C_j-A_j$. \texttt{makespan\_p95\_s} is
$Q_{0.95}(\{T_j:j\in J_c\})$, {\leavevmode the 95th percentile of completed-job latencies}. Throughput is
$\Theta=|J_c|/T_{\mathrm{run}}$. Completion ratio
$\rho_{\mathrm{done}}=|J_c|/|J|$ marks unfinished workload-scaling configurations as
capacity/horizon signals. For a resource class $r$, the active-wall critical
path fraction (CPF) is $\phi_r=B^{\mathrm{wall}}_r/T_{\mathrm{run}}$, reported
in the \texttt{cpf\_$r$} or active-utilization fields. The bottleneck class is
$\arg\max_r \phi_r$. Capacity-normalized utilization is
$U_r=B_r/(T_{\mathrm{run}}\mathrm{cap}_r)$ and is used only where the results
explicitly report utilization curves. {\leavevmode $B^{\mathrm{wall}}_r$ is elapsed time with at least one active instance of class $r$, $B_r$ is accumulated busy capacity-time, and $\mathrm{cap}_r$ is total class capacity.}

{\leavevmode A mixed bundle contains jobs from multiple workload families sharing one HCU. ``Jobs per workload'' specifies the number submitted for each workload family in that bundle. A stage is one resource-consuming step within a job, so completing a stage does not imply completing the job. Run makespan and the p95 of per-job completion times summarize different quantities, as defined above.}

{\leavevmode Supplementary deployment, mixed-workload, and sensitivity results: project documentation~\cite{hybridqc_project}.}

\subsection{V1: Hardware Grounding Is Accurate Enough for System-Level and Multi-Stage Simulation}
Table~\ref{tab:validation_error} reports compact validation anchors. For
D-Wave, the best Advantage 2 training fit reaches 3.92\% MAPE and
$R^2=0.981$ for QPU access time. The current workload-validation fit for
Advantage 1 reaches 8.04\% MAPE and $R^2=0.994$. For IBM, Fez reaches
5.26\% MAPE on its backend campaign, Kingston reaches 16.45\% MAPE with the
transpiled augmented model, and Marrakesh reaches 19.01\% MAPE. These are
multi-stage service fits, using the metrics defined in
Section~\ref{sec:evaluation}. They are not fits to end-to-end cloud wall time:
cloud wall time includes queueing, provider orchestration, client overhead, and
network effects that vary across submissions and are modeled separately from
physical QPU occupancy.

The result supports the first validation claim: hardware measurements can
ground the QA and DQC occupancy models used by HybridQC. It also shows why backend
profiles must be explicit. Kingston, Marrakesh, and Fez differ enough that a
single global IBM coefficient set would hide backend-specific behavior. The
same is true for D-Wave Advantage 1 versus Advantage 2: the profiles share a
generic QA structure, but the fitted parameters and hardware metadata should
remain backend-specific.
{\leavevmode The supporting data~\cite{hybridqc_supplementary_data} provides the complete calibration and workload-validation measurements, fitted parameters, fit diagnostics, and timing breakdowns.}

\begin{table}[t]
  \centering
  \footnotesize
  \setlength{\tabcolsep}{2.4pt}
  \renewcommand{\arraystretch}{1.08}
  \begin{tabular}{@{}p{0.23\linewidth}p{0.33\linewidth}rrr@{}}
    \toprule
    Backend & Timing target & $n$ & MAPE (\%) & $R^2$ \\
    \midrule
    D-Wave Adv.-1 & QPU access time & 240 & 8.04 & 0.994 \\
    D-Wave Adv.-2 & QPU access time & 263 & 3.92 & 0.981 \\
    IBM Fez & quantum seconds & 7 & 5.26 & 0.978 \\
    IBM Kingston & quantum seconds & 80 & 16.45 & 0.978 \\
    IBM Marrakesh & quantum seconds & 21 & 19.01 & 0.967 \\
    \bottomrule
  \end{tabular}
  \caption{Hardware-grounding anchors. The timing
  target is the physical occupancy quantity used by the simulator profile.
  Counts give the number of usable hardware measurements after filtering failed,
  unsupported, or incomplete runs. D-Wave campaigns therefore have more measurements
  than the IBM seed profiles, which were limited by runtime-credit and backend
  availability. {\leavevmode Complete measurements and fit diagnostics} are in the {\leavevmode supporting data~\cite{hybridqc_supplementary_data}}.}
  \label{tab:validation_error}
\Description{Five backend rows give the timing target, measurement count, mean absolute percentage error, and coefficient of determination. Two rows cover D-Wave quantum processing unit access time and three cover IBM quantum seconds.}
\end{table}

\paragraph{Insight.}
The simulator should not use provider wall time as physical QPU occupancy.
For D-Wave, QPU access time is the right resource-locking target. Wall time is
better modeled as cloud/client overhead. For IBM, the right DQC-service target
is IBM-reported quantum seconds, while queue and orchestration fields remain
deployment overheads. This separation is central to using current cloud data
to forecast future tightly integrated systems.

\subsection{E1: Balanced Topologies Matter More Than Isolated QPU Scaling}

\begin{table}[t]
\centering
\begingroup
\fontsize{8}{10}\selectfont
\setlength{\tabcolsep}{4pt}
\begin{tabular}{@{}lrrrr@{}}
\toprule
\multicolumn{5}{c}{Mean p95 completion time $\pm$ SD (s)} \\
Resource & x1 & x2 & x3 & x4 \\
\midrule
QA & 610.5 $\pm$ 814.2 & 610.5 $\pm$ 814.2 & 610.5 $\pm$ 814.2 & 610.5 $\pm$ 814.2 \\
DQC & 610.5 $\pm$ 814.2 & 533.6 $\pm$ 767.4 & 503.6 $\pm$ 759.6 & 467.5 $\pm$ 749.2 \\
QA ctrl. & 610.5 $\pm$ 814.2 & 611.3 $\pm$ 818.1 & 611.2 $\pm$ 816.6 & 608.6 $\pm$ 817.6 \\
DQC ctrl. & 610.5 $\pm$ 814.2 & 606.5 $\pm$ 800.9 & 601.3 $\pm$ 792.4 & 587.1 $\pm$ 761.7 \\
\midrule
\multicolumn{5}{c}{Mean throughput (jobs/s)} \\
Resource & x1 & x2 & x3 & x4 \\
\midrule
QA & 2.58 & 2.58 & 2.58 & 2.58 \\
DQC & 2.58 & 2.61 & 2.61 & 2.62 \\
QA ctrl. & 2.58 & 4.50 & 5.74 & 6.83 \\
DQC ctrl. & 2.58 & 2.58 & 2.58 & 2.58 \\
\bottomrule
\end{tabular}
\caption{{\leavevmode Replica scaling under HOBA. Columns give the total instances of the named resource; all other resource counts remain fixed. Completion times are the mean $\pm$ sample SD of per-run p95 values across 11 workload families, each with 30 completed jobs. Throughput is the mean across the same families. SD describes between-family variation.}}
\label{tab:replica_payoff_main}
\Description{Two blocks report completion time and throughput for one through four instances of each quantum device or controller resource. Completion time entries include the mean and standard deviation across workload families; throughput entries give the mean.}
\endgroup
\end{table}

\begin{figure}[t]
  \centering
  \includegraphics[width=\linewidth]{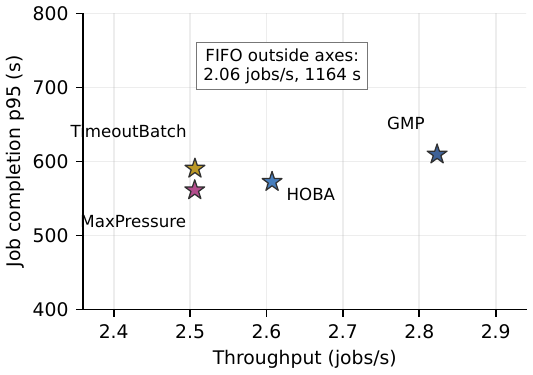}
  \caption{Policy frontier for p95 job-completion time and throughput,
  regenerated from the current Rust experiment outputs. {\leavevmode The policy panel
  focuses on 400--800 s; the FIFO point is reported outside the axes at its
  original coordinates.}}
  \label{fig:e1_policy_tradeoff_main}
  \Description{A scatter plot compares scheduling policies using throughput on the horizontal axis and the 95th percentile of job-completion time on the vertical axis. Each labeled star represents one policy. GMP has the highest throughput among the plotted policies. FIFO falls outside the displayed vertical range and is reported in a text box at 2.06 jobs per second and 1164 seconds. The other policies trade throughput against tail completion time.}
\end{figure}

\begin{figure}[t]
  \centering
  \includegraphics[width=\linewidth]{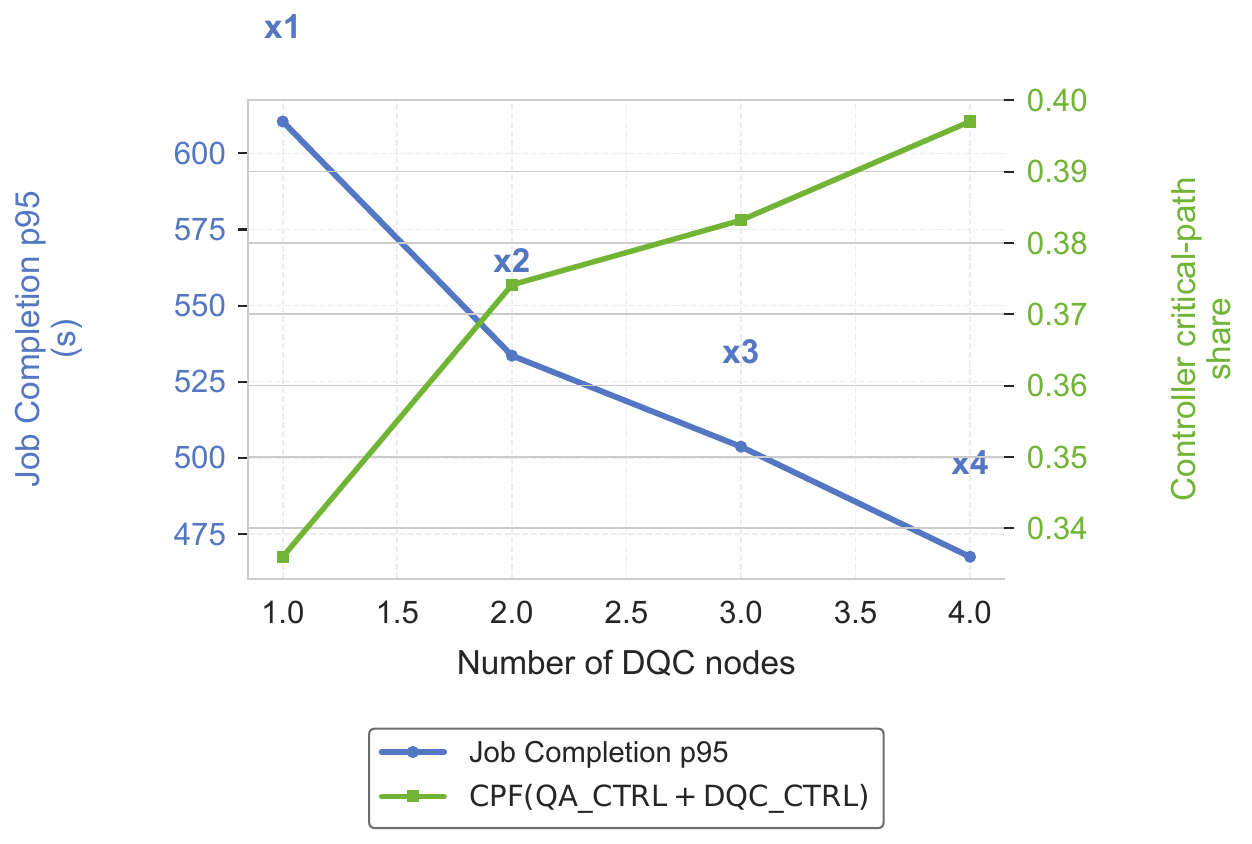}
  \caption{DQC bottleneck migration under replica scaling. Completion time
  improves as DQC capacity increases, but controller CPF grows, exposing the
  next serialized component. CPF denotes the implementation's active-wall
  bottleneck diagnostic for the corresponding resource class. {\leavevmode The left axis reports p95 job-completion time; the right axis reports the combined QA- and DQC-controller CPF as the DQC instance count increases from one to four.}}
  \label{fig:e2_dqc_migration_main}
  \Description{A dual-axis line plot increases the number of digital-quantum devices from one to four. The blue curve for the 95th percentile of job-completion time falls, while the green curve for the combined quantum-annealing and digital-quantum controller critical-path fraction rises. The opposing trends show that additional quantum-device capacity reduces completion time while increasing the relative importance of the controller path.}
\end{figure}

\begin{figure}[t]
  \centering
  \includegraphics[width=\linewidth]{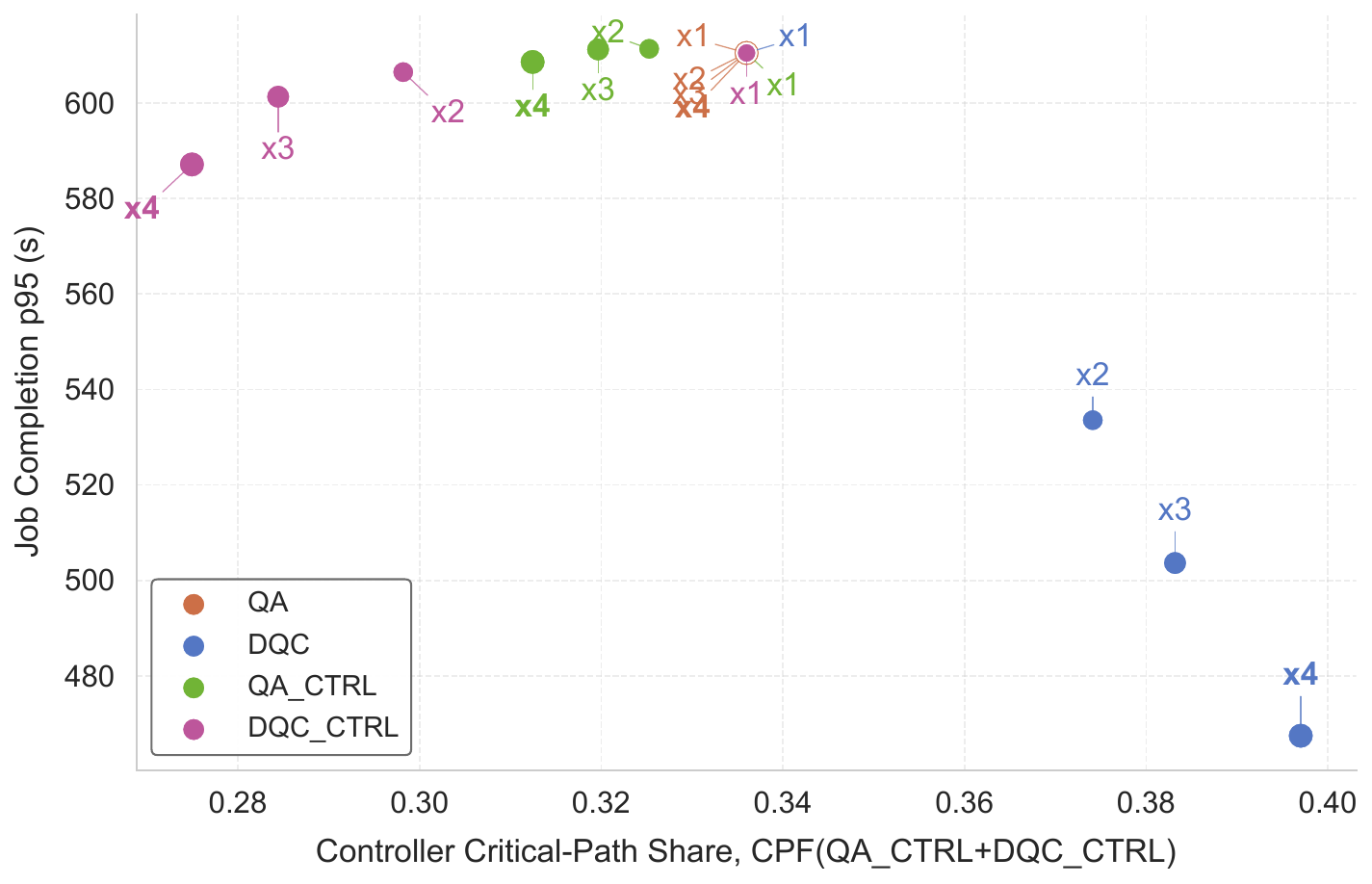}
  \caption{Bottleneck migration as resource replicas change. The plot separates
  the completion-time payoff from the controller-share movement so this
  diagnostic can be inspected independently from the main-text topology
  summary. {\leavevmode Colors identify the replicated resource class, and x1--x4 label its total instance count. The horizontal axis is the combined controller CPF and the vertical axis is p95 job-completion time.}}
  \label{fig:e2_bottleneck_migration_main}
  \Description{A scatter plot places the combined controller critical-path fraction on the horizontal axis and the 95th percentile of job-completion time on the vertical axis. Four colors identify separate replica sweeps for quantum-annealing devices, digital-quantum devices, and their controllers. Labels x1 through x4 identify total instance counts. Digital-quantum device replication moves the points toward higher controller fraction and lower completion time. The other sweeps occupy distinct, more tightly clustered regions.}
\end{figure}

{\leavevmode Table~\ref{tab:replica_payoff_main}} shows why HybridQC~models the whole HCU
rather than only QPU count. {\leavevmode The sweep uses the 11 workload families summarized in Table~\ref{tab:workload_families}, each run separately as described in Section~\ref{sec:evaluation}.} {\leavevmode Its rows separately report QA-device, DQC-device, QA-controller, and DQC-controller replication.} The labels x1--x4 denote the total number of instances of the swept
resource class, where x1 is the baseline and x4 is the four-replica variant.
{\leavevmode Average throughput remains 2.58 jobs/s at each replica count in the QA-device and DQC-controller sweeps. For these workloads, the replicated resource is not on the active bottleneck path, so adding replicas leaves average throughput unchanged.}
{\leavevmode Figures~\ref{fig:e2_dqc_migration_main} and~\ref{fig:e2_bottleneck_migration_main} show the active-time diagnostics that explain this bottleneck migration.}

The first topology result is that the largest payoff does not always come from
replicating the quantum device itself. Moving from one to four QA-controller
replicas improves throughput by 164\% while changing p95 completion time by
only $-2$ s. Replicating DQC devices alone improves throughput by roughly 1\%,
but controller active time rises as DQC replicas are added. Thus
{\leavevmode Table~\ref{tab:replica_payoff_main}} shows broad replica payoff, while the
{\leavevmode plots in Figures~\ref{fig:e2_dqc_migration_main} and~\ref{fig:e2_bottleneck_migration_main}} show why QPU scaling helps most when the matching
controller path scales with it. The balanced 10$\times$ topology remains the
best variant, but its speedup is 2.19--3.42$\times$, not 10$\times$.

{\leavevmode Complementary pair sweeps jointly vary a QPU count and its matching controller count, reporting means across the same 11 workload families under the HOBA policy.} QA-controller replication gives the strongest throughput improvement:
average throughput increases from 2.85 to 9.75 jobs/s, a 3.43$\times$ gain,
while p95 completion time changes little. DQC-controller replication gives the
stronger tail-latency improvement: p95 completion time falls from 7601 s to
6348 s, about 16.5\%, while throughput changes by only about 1\%. Thus the
best topology extension depends on the objective. A QPU-only model would miss
this split because the critical resource is often the controller path rather
than the QPU itself. {\leavevmode The supporting data~\cite{hybridqc_supplementary_data} keeps} the corresponding pair-sweep numerical summary for auditability.

\paragraph{Insight.}
Adding only quantum capacity can be the wrong investment: DQC-heavy workloads
can remain controller- or reconstruction-limited, QA-heavy workloads can remain
embedding- or transfer-limited, and mixed workloads can expose both.
{\leavevmode The supporting policy/topology summary~\cite{hybridqc_supplementary_data} preserves} the key unbalanced
variants that do not fit in the main text.

\subsection{E2: Scheduling Policy Changes Throughput and Bottleneck Migration}

Scheduling changes the mixed-workload result because policies use different signals when several ready stages compete for the same HCU resource. Section~\ref{sec:policies} defines the policies and their dispatch rules.

{\leavevmode Figure~\ref{fig:e1_policy_tradeoff_main} averages per-run throughput and p95 completion time across 11 workload families and nine stage configurations, with 30 jobs per workload family run. Within each configuration, the workload and topology are fixed while the scheduling policy changes.} FIFO has the lowest throughput and highest tail latency in the aggregate policy sweep, while GMP has the highest throughput. {\leavevmode The active-time breakdown in the supporting results~\cite{hybridqc_project} shows how the policies shift resource pressure}:
different policies shift bottleneck pressure across QA, DQC, controller,
transfer, and classical components rather than merely changing a single queue
order.
{\leavevmode A separate stress study co-schedules six workload families on one HCU with compilation, cutting, and QEM stages enabled, comparing FIFO, HOBA, and Max-Pressure at 5, 10, and 20 jobs per workload family. At 20 jobs per workload family (120 jobs total)}, HOBA finishes in 3812.66 s, Max-Pressure in
4876.15 s, and FIFO in 6862.32 s. Thus HOBA is 1.80$\times$ faster than FIFO
for this stress case. The best policy is not universal: GMP is best for bin
packing, HOBA for portfolio, and Max-Pressure for PES scan and the {\leavevmode mixed workload with 10 jobs per workload family}.

\paragraph{Insight.}
Policy value depends on workload composition. FIFO is useful as a baseline, but
pressure-aware and bottleneck-aware policies can improve throughput when they
see the right signal and can over-prioritize the wrong queue when bottlenecks
shift. This motivates simulating policy and topology together.
{\leavevmode Section~\ref{sec:policies} gives the policy scoring signals, and supporting data~\cite{hybridqc_supplementary_data} retains the policy-by-workload comparisons behind this summary.}

\subsection{S1: Workload Scaling Is Multi-Axis}

\begin{figure*}[!t]
  \centering
  \begin{minipage}[t]{0.32\textwidth}
    \centering
    \textbf{(a)}
    \includegraphics[width=\linewidth]{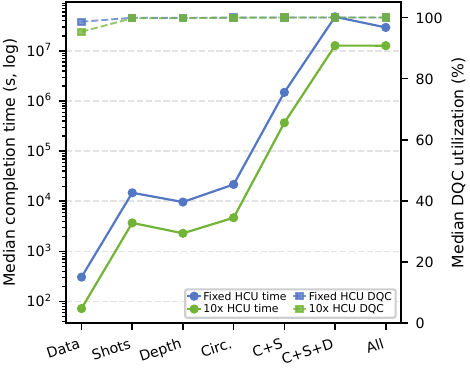}
  \end{minipage}\hfill
  \begin{minipage}[t]{0.32\textwidth}
    \centering
    \textbf{(b)}
    \includegraphics[width=\linewidth]{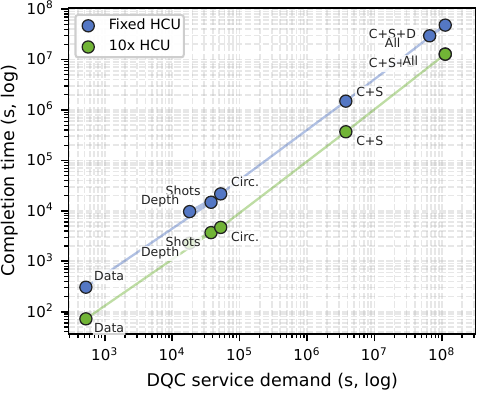}
  \end{minipage}\hfill
  \begin{minipage}[t]{0.32\textwidth}
    \centering
    \textbf{(c)}
    \includegraphics[width=\linewidth]{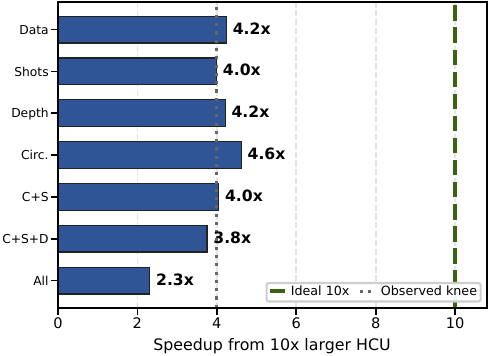}
  \end{minipage}
  \caption{Workload-scaling analysis. (a) 100$\times$ scaling by growth axis.
  Solid curves use the left log-runtime axis and dashed curves use the right
  utilization axis. (b) Bottleneck migration under fixed and 10$\times$ larger
  HCUs. (c) Median-runtime speedup from moving 100$\times$ workloads to the
  10$\times$ larger HCU. The dashed line marks ideal 10$\times$ improvement.
  Abbreviations: data = payload/memory, shots = reads/shots, c = circuits,
  c+s = circuits+shots, c+s+d = circuits+shots+depth, all = all fields.}
  \label{fig:workload_scaling_main}
  \label{fig:future_axis_scaling_main}
  \label{fig:future_bottleneck_migration_main}
  \label{fig:future_compromise_speedup_main}
  \Description{Three panels compare workload-growth axes. Panel a shows that increasing data alone causes much less runtime growth than jointly increasing circuit count, shots, and depth. Panel b compares resource bottlenecks for the fixed machine and a machine with ten times its resources. Panel c shows that the larger machine gives different runtime speedups for different growth axes, with every bar below the dashed line marking an ideal tenfold speedup.}
\end{figure*}

S1 scales the workload families in Table~\ref{tab:workload_families}, with
full field ranges in {\leavevmode Table~\ref{tab:workload_parameter_ranges}}: QA
optimization, DQC circuit workloads, cross-modality QA+DQC workflows, and mixed
bundles sharing the same HCU. The largest current DQC configurations in {\leavevmode that table} have
up to 72 logical qubits and depth 2,000, so 10$\times$ and 100$\times$
all-field scaling stress future systems with hundreds to thousands of logical
qubits and depths up to 200,000 rather than hardware that exists today.

The strongest scaling result is that growth is not one-dimensional. At
100$\times$ data-only scaling on the fixed machine, median runtime is about
306 s. At 100$\times$ circuits-shots-depth scaling, it reaches about
$4.806\times 10^7$ s on the fixed machine and $1.280\times 10^7$ s on the
10$\times$ larger machine. Figure~\ref{fig:workload_scaling_main}(a) separates
these axes, while Figure~\ref{fig:workload_scaling_main}(b) shows how those
axes move the bottleneck. QA-only workloads grow much more slowly than DQC-only
and mixed QA+DQC workloads under multi-axis workload scaling. The large values
are not a transfer artifact. They come from DQC growth where circuit count,
shots, and depth multiply. Infeasible 100$\times$ all-fields configurations are reported
separately as capacity failures, not as completed runtime ratios.

Figure~\ref{fig:workload_scaling_main}(b) explains why the larger HCU does not
recover ideal scaling: circuits, shots, and depth move points along the same
DQC-dominated trajectory, so balanced capacity lowers runtime without changing
the root cause of saturation.

Figure~\ref{fig:workload_scaling_main}(c) quantifies the compromise. At
100$\times$ workload scale, the 10$\times$ larger HCU improves median runtime
by 4.24$\times$ for data-only growth, 4.62$\times$ for circuit-only growth,
4.05$\times$ for circuits-and-shots growth, 3.75$\times$ for
circuits-shots-depth growth, and 2.31$\times$ for all-fields growth. The
larger topology helps most when the scaled workload exposes parallel work, but
it does not produce linear speedup when multiple DQC execution axes grow
together.

\paragraph{Insight.}
Workload-scaling studies should report the scaling axis, not only the scale
factor. Data-only scaling mostly stresses transfer, memory, and preprocessing,
whereas circuits-shots-depth scaling multiplies submitted circuits, shots per
circuit, and circuit depth. HybridQC~turns that ambiguity into an experiment by
scaling workload fields separately and recording which resources saturate.
{\leavevmode Supporting data~\cite{hybridqc_supplementary_data} retains the representative completed 100$\times$ configurations and the unfinished fixed-HCU capacity-boundary configurations.} 

\section{Discussion and Limitations}
\label{sec:discussion}

HybridQC~preserves stage graphs, finite queues, payloads, controllers, and transfer routes when staged hybrid workloads contend for a proposed machine. Hardware measurements are used to make the service models credible at today's scale, while the simulator uses those calibrated models to ask architecture questions that cannot yet be measured directly, such as how controller, link, and scheduling bottlenecks move in future local-accelerator HCUs.

The main limitations are scope and extrapolation. HybridQC~is not a circuit-state
simulator, annealing solver, quantum-control compiler, or provider Runtime
replacement. It consumes circuit/QUBO properties, backend measurements,
topology definitions, and stage graphs. The fitted QA/DQC service models
reflect the hardware data collected here, but future backends, provider queues,
circuit families, embedding outcomes, and algorithmic accuracy require new
calibration. Workload scaling is therefore a controlled stress test:
capacity-infeasible {\leavevmode configurations} are architecture warnings under the configured
memory, transfer, and resource constraints, not universal predictions.

{\leavevmode The evaluated claims concern runtime, resource contention, and feasibility under the configured capacities. The simulator does not evolve a quantum state or establish the algorithmic correctness of a workload. The QEC protection stages defined in Section~\ref{sec:model} are disabled in these studies, and device reliability is a separate modeling question.}

\section{{\leavevmode Supporting Material and Availability}}
\label{sec:availability}
{\leavevmode Supporting data is available online~\cite{hybridqc_supplementary_data}. Extended methods, figures, tutorials, and configuration examples are in the project repository~\cite{hybridqc_project}. Source-code release is pending institutional authorization from Fordham University.}

\section{Conclusion}
\label{sec:conclusion}

HybridQC~provides a hardware-grounded simulator for tightly integrated hybrid
quantum-classical systems. By validating service models on live IBM and D-Wave
measurements and applying them through a topology-aware discrete-event engine,
it lets researchers evaluate full hybrid workloads and forecast bottlenecks in
machines that do not yet exist. The evaluation shows that
backend-specific profiles can ground system-level QA and DQC occupancy, that
balanced resource scaling is more effective than isolated QPU scaling, that
scheduling policy changes mixed-workload throughput, and that cloud wall time
should be separated from physical resource occupancy. It also shows that future
workload scaling must be described by axes: data-only scaling is mild, while
joint growth in circuit count, shots, and depth can saturate DQC resources by
orders of magnitude. HybridQC~therefore offers a practical path for validating
today's hybrid workloads and forecasting tomorrow's architecture bottlenecks.

\clearpage
\bibliographystyle{ACM-Reference-Format}
\bibliography{paper_sections/references}

\end{document}